\documentclass[aip,reprint]{revtex4-1}
\pdfoutput=1

\usepackage{graphicx}
\usepackage{amsmath}
\usepackage{amssymb}

\begin{document}


\title{Transient localized wave patterns and their application to migraine} 




\author{Markus A. Dahlem}
\email[]{dahlem@physik.tu-berlin.de}
\homepage[]{https://sites.google.com/site/markusadahlem/}

\affiliation{Institute for Physics, Humboldt Universit\"at zu Berlin, Berlin, Germany}
\affiliation{Institute for Theoretical Physics, Technische Universit\"at Berlin, Berlin, Germany}

\author{Thomas M. Isele}
\affiliation{Institute for Theoretical Physics, Technische Universit\"at Berlin, Berlin, Germany}


\date{\today}

\begin{abstract}
Transient dynamics is  pervasive in the human brain and  poses challenging
problems both in mathematical tractability and clinical observability. We
investigate statistical properties of transient cortical wave patterns with
characteristic forms (shape, size, duration) in a canonical reaction-diffusion
model with mean field inhibition.  The patterns are formed by a ghost near a
saddle-node bifurcation in which a stable traveling wave (node) collides with
its critical nucleation mass (saddle). Similar patterns have been observed with
fMRI in migraine. Our results support the controversial idea that waves of
cortical spreading depression (SD) have a causal relationship with the headache
phase in migraine and therefore occur not only in migraine with aura (MA) but
also in migraine without aura (MO), i.\,e., in the two major migraine subforms.
We suggest a congruence between the prevalence of  MO and MA with the
statistical properties of the traveling waves' forms, according to which (i) activation of
nociceptive mechanisms relevant for headache is dependent upon a sufficiently
large instantaneous affected cortical area anti-correlated to both SD duration
and total affected cortical area such that headache would be less severe in MA
than in MO (ii) the incidence of MA is reflected in the distance to the
saddle-node bifurcation, and (iii) the contested notion of MO attacks with
silent aura is resolved. We briefly discuss model-based control and means by
which neuromodulation techniques may affect pathways of pain formation.
\end{abstract}


\maketitle 




\section*{Author Summary}

The key to the genesis of migraine with aura is a traveling wave phenomenon
called cortical spreading depression (SD). Migraine is characterized by
recurrent episodes, the aura phase usually lasts only about 30$\,$min. Thus, SD
is a transient state. During its course, SD massively perturbs the brain's
ion homoeostasis. We resolve the puzzling problem why SD does not engulf all
of the densely packed excitable neurons by suggesting a well-established
pattern formation mechanism of long-range inhibitory feedback. Furthermore, we
use cortical feature maps to create plausible initial conditions as
perturbations of the homogeneous state.  The statistics of occurrences of the
different classes has the potential to reproduce epidemiological statistics of
different diagnostic forms of migraine such as migraine with or without
aura and provides simple answers to some very controversially discussed
current questions in migraine research.

\section*{Introduction}
 
The undoubtedly most fundamental example of transient dynamics is the dynamical
phenomenon of excitability, that is, all-or-none behavior.  Shortly after
transient response properties of excitable membranes were classified into two
classes\cite{HOD48}, it was also explained in a detailed mathematical model how
excitability emerges from electrophysiological properties of such membranes in
the ground-breaking work by Hodgkin and Huxley\cite{HOD52}.  Two features are
central and are by no means exclusive to biological membranes but shared by all
excitable elements.  Firstly, the inevitable threshold in any all-or-none
behavior requires nonlinear dynamics.  Secondly, the transient response of the
system to a super-threshold stimulation eventually has to lead back to a
globally stable steady state after some large phase space excursion.  This
indicates global dynamics, that is, dynamics involving not only fixed points
and their local bifurcations but larger invariant sets, for instance periodic
orbits that collide with fixed points. An excitable element is in some sense
the washed-up brother of the relaxation oscillator: when the threshold
vanishes, a single excitable element usually becomes a simpler behaved---and much
longer known---relaxation oscillator\cite{POL26}.  

In this study, we propose a model for wave patterns with a characteristic
shape, size, and duration. These waves are transient responses to confined,
spatially structured perturbations of the homogeneous steady state. The
homogeneous steady state is globally stable in our proposed reaction-diffusion
model because we also introduce an effective inhibitory mean field feedback
control. This leads to a new type of local excitability in a spatially extended
medium involving disappearing traveling wave solutions as larger invariant
sets. Both, the model and the initial conditions are motivated by the
pathophysiology of migraine and clinical observations
\cite{DAH00a,HAD01,DAH08d} and the results are applied to some currently
controversial topics \cite{AYA10,EIK10,AKE11}. 

We will briefly introduce concepts of excitable elements and excitable media in
two-variable reaction-diffusion systems and also the idea of an additional
long-range inhibitory feedback that is studied in various other systems outside
the neurosciences and also in neural field models. While we also briefly
introduce migraine, the view of migraine as a dynamical disease is more
elaborated in the discussion.

The original conductance-based membrane model from Hodgkin and Huxley, and the
more refined versions to date, contain many variables, but fortunately this is
not essential for excitable elements. In fact, it turned out that the two
classes of excitability are actually amenable to direct analysis in a
two-dimensional phase plane by identifying fast and slow processes in the
conductance-based model  and grouping these into dynamics of just two lump
variables\cite{BON53,FIT69}. Using such a geometrical approach and partly analytical
theory, the original empirical classification of excitability was further
pursued with bifurcation analysis\cite{RIN89}, explaining class I by
identifying its threshold as a stable manifold of a saddle point on an
invariant cycle and the one of class II as a trajectory from which nearby
trajectories diverge sharply (called a canard trajectory). Extensions to these
principal mechanisms involve codimension 2 bifurcations and lead also to
bursting in three-variable models, which have been investigated in great
detail\cite{IZH00a}. However, the two-variable models of a fast activator and
slow inhibitor and their phase portraits of class I and II became qualitative
prototypes for excitable systems in various biological\cite{KEE98},
chemical\cite{KAP95a} and physical contexts\cite{SCH87}. 

\begin{figure}[t]
\includegraphics[width=0.95\columnwidth]{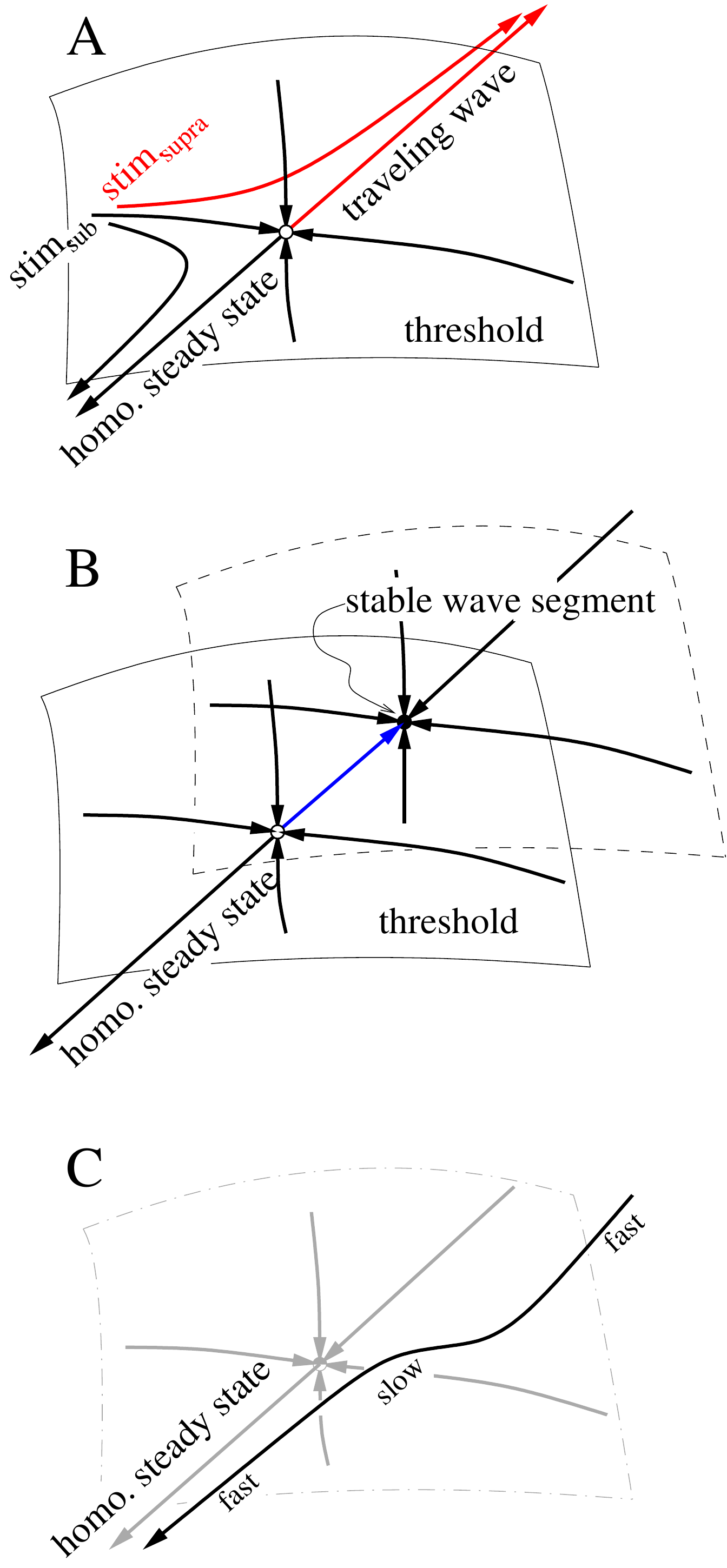}
\caption{\label{fig:phase_space_sketch}   Schematic sketch of the phase space
of (A) the uncontrolled system, (B) the system with mean field control adjusted
such that the nucleation solution is stabilized and (C) the system with mean field
control and control parameters such that the ghost of the saddle-node
bifurcation is still influencing the dynamics.}
\end{figure}

Distinct from these excitable systems are those that are spatially extended
systems, called excitable media. (We use the word ''system'' as a general term,
''medium'' only for spatially extended systems, and ''element'' for point-like
systems.) Already the original work by Hodgkin and Huxley\cite{HOD52} described
extended tube-like membranes (axons) and introduced the cable equation as a
parabolic partial differential equation, which is in the same class as the
diffusion equation.  Even in reaction-diffusion media with infinite-dimensional
phase space, we can again apply geometrical approaches, simply because
excitable media are not defined by---in contrast to excitable elements---transient
dynamics but traveling wave solutions.  The quiescent state is the non-excited
homogeneous steady state. And like the quiescent state, excited states of a
medium are usually stationary states in some appropriate comoving frame with
$\xi=x-ct$. Furthermore, the threshold is related to an unstable stationary
state, the critical nucleation solution, usually in another comoving frame
including $c=0$. The existence of the nucleation solution is a simple
consequence of multistability, see Fig.~\ref{fig:phase_space_sketch}A, but note
that even monostable excitable elements have similar unstable stationary states
in class I.

Central to our approach is a distinct subexcitable medium in which  localized
traveling waves occur only transiently.  Reaction-diffusion waves would engulf
all of the medium, if formed in a two-variable system with only one activator
and one inhibitor with the system's parameters in the appropriate regime.  In
contrast, localized traveling waves indicate a demand-controlled excitability.
Similar ideas to obtain localized traveling, though not transient, waves have
been introduced in various contexts, for instance an integral negative feedback
or a third, fast diffusing inhibitory component for moving spots in
semiconductor materials, gas discharge phenomena, and chemical
systems\cite{OHT89a,KRI94,SCH97a,SAK02}.  Furthermore, in neural field
models\cite{BRE12}, localized two-dimensional bumps are studied
\cite{LU11a,BRE11} in integrodifferential equations (without diffusion) in the
context, for example, of memory formation\cite{KIL13}. Localized structures have also been
discussed in the context of cortical spreading depression (SD) in migraine
before, in particular a model with narrowly tuned parameters that shows
transient waves \cite{DAH03a,DAH07a,DAH08d} and a model with mean field
feedback control that allows for localized waves \cite{DAH09a}. But it is for
the first time now that a model is presented in which wave phenomena occur that
are both localized and transient, so that a variety of new questions that are
controversially discussed in migraine research\cite{AYA10,AKE11,FIO11b,LEV12}
can be adressed.  

\begin{figure*} \begin{center}
\fbox{\includegraphics[width=\textwidth]{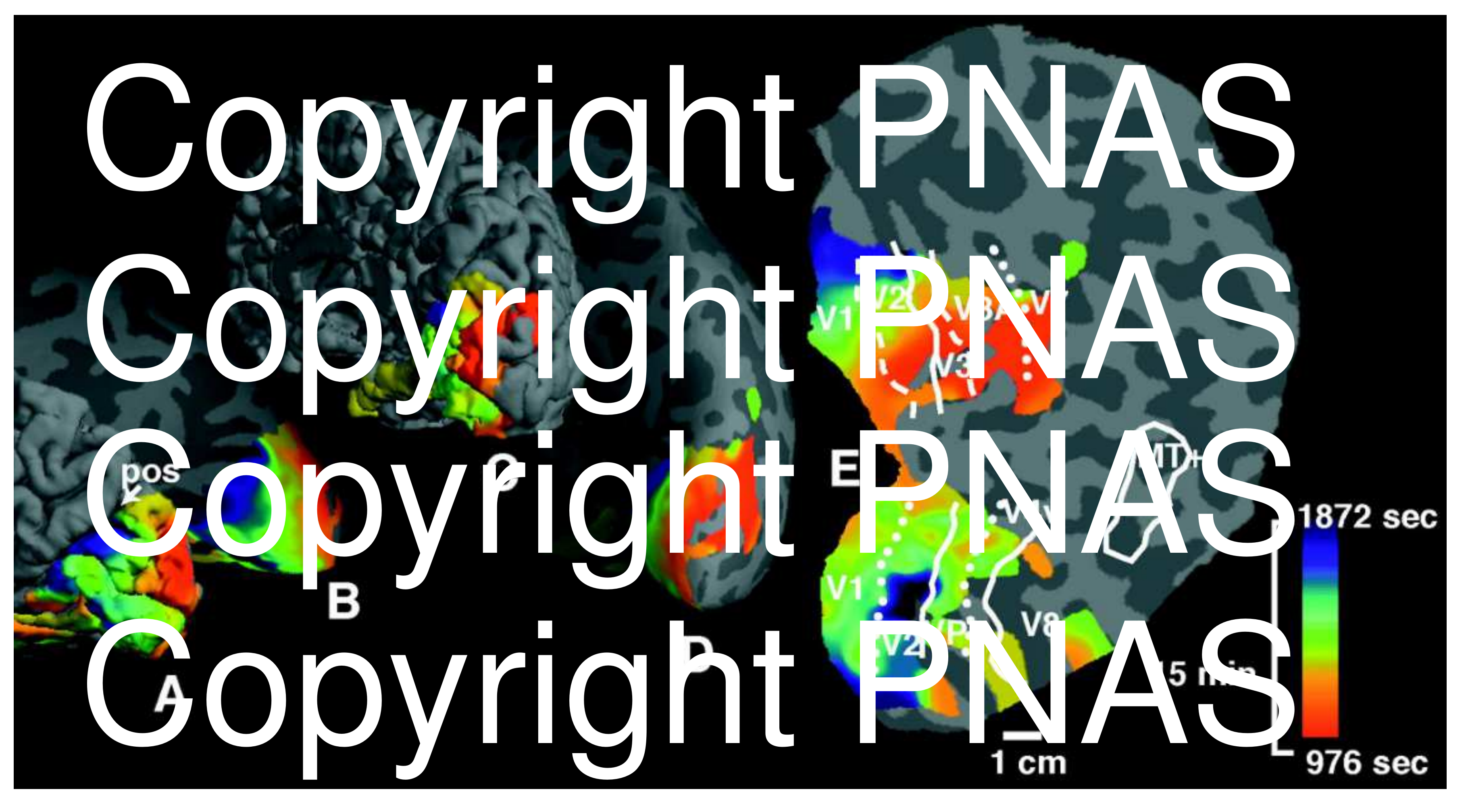}} \end{center}
\caption{\label{fig:had01_fig4} Source localization of the magnetic resonance
(MR) data signal of SD (from \cite{HAD01}). Color code: time from onset,
locations showing the first MR signals of SD are coded in red, later times are
coded by green and blue (see color scale to the right). Signals from the first
975 seconds were not recorded because the migraine attack was triggered outside
the MR imaging facility. (A) The data on folded right posterior pole
hemispheric cortex; (B) the same data on inflated cortical surface; (C and D)
the same data shown on the entire hemisphere from posterior-medial view
(oblique forward facing), folded and inflated, respectively. As described in
the original study \cite{HAD01}, MR data were not acquired from the extreme
posterior tip of the occipital pole (rearmost portion). (E) A fully flattened
view of the cortical surface.  The aura-related changes are localized wave
segments. Note that in the flattened cortex was cut along the steep sulcus
calcarine to avoid large area distortions induced by the flattening process.
The colored border to the left is the cut edge that should be considered being
connected such that the color match up as seen in (B and D).  (Copyright
permission from authors granted, from PNAS is requested.) } \end{figure*}

Migraine is characterized by recurrent episodes of head pain, often throbbing
and unilateral. In migraine without aura (MO), attacks are usually associated
with nausea, vomiting, or sensitivity to light, sound, or movement
\cite{GOA02}. Migraine with aura (MA) involve, in addition but also rarely
exclusively, neurologic symptoms (aura) that are associated with waves of
cortical SD \cite{OLE81,HAD01}. SD is a reaction-diffusion process, although,
clearly, the originally proposed mechanism of a simple one-variable model for
SD front propagation triggered by diffusion of elevated extracellular potassium
with bistable kinetics, to date known as Hodgkin-Huxley-Grafstein
model\cite{GRA63}, does not capture the complex chain of involved
reactions\cite{SOM01,STR05,HER05}.  However, migraine aura symptoms manifest
themselves on a macroscopic scale over several minutes up to one hour and
extend over several centimeters when mapped onto the corresponding cortical
areas\cite{HAD01,VIN07}, see Fig.~\ref{fig:had01_fig4}.  In this study, we are
interested in these clinically relevant properties of migraine with aura.   To
this end, we exploit the concept of nucleation, growth, and subsequent
shrinking of SD in a canonical reaction-diffusion model of activator-inhibitor
type with mean field feedback control.

\section*{Results} 
\label{sec:results}

We first present a model constructed by adding a mean field inhibitory feedback to a well known
reaction-diffusion system, then we present
the statistical properties of the transient behavior this model is exhibiting.

\subsection*{Before infinity and beyond by mean field inhibitory feedback control}

The diversity of the behavior of traveling waves in two spatial dimensions was
studied in canonical models (see Discussion) depending on the two generic
parameters $\beta$ and $\varepsilon$ in Eqs.\,(\ref{eq:fhn1})-(\ref{eq:fhn2}),
which determine the parameter plane of excitability \cite{WIN91}.
Eqs.\,(\ref{eq:fhn1})-(\ref{eq:fhn2}) determine an excitable medium without
feedback control. In those media, patterns of discontinuous (open ends)
spiral-shaped waves are used to probe excitability and these patterns are
closely related to the discontinuous, localized transient waves  we propose in
our model.

We make use of the fact that at a low critical excitability, called the rotor
boundary $\partial R_{\infty}$, spiral waves do not curl-in anymore but become
half plane waves\cite{MIK91,HAK99}.  Beyond the rotor boundary lies the
subexcitable regime in which discontinuous waves start to retract at their open
ends and any discontinuous wave is transient and will eventually disappear.
The border $\partial R_{\infty}$ marks a saddle-node bifurcation at which
discontinuous waves collide with their corresponding nucleation solution.  This
leads to the key idea of our model. A linear mean field feedback control moves
this saddle-node bifurcation towards distinct localized wave segments with a
characteristic form (shape, size) and behind this bifurcation these waves
become transient objects (see Fig.~\ref{fig:phase_space_sketch}). 

Before we introduce the effect of mean field feedback control, we have to
consider the behavior of continuous waves (closed waves fronts without open
ends) when the excitability is decreased. This will be important if we want to
understand the fate of any solution, discontinuous or not, under mean field
feedback control.  Unbroken plane waves propagate persistently even if the
parameters are chosen in the subexcitable regime until the propagation boundary
$\partial P$ is reached.  At this border, the medium's excitability becomes too
weak for continuous plane waves to propagate persistently.  The border
$\partial P$ in parameter space indicates again a saddle-node bifurcation at
which a planar traveling wave solution collides with its corresponding
nucleation solution. Note, that the planar wave is essentially a pulse solution
in 1D and the nucleation solution in 1D is called the slow wave \cite{KRU97}.

\begin{figure*}[t]
  \begin{center}
    \fbox{\includegraphics[width=1.0\textwidth]{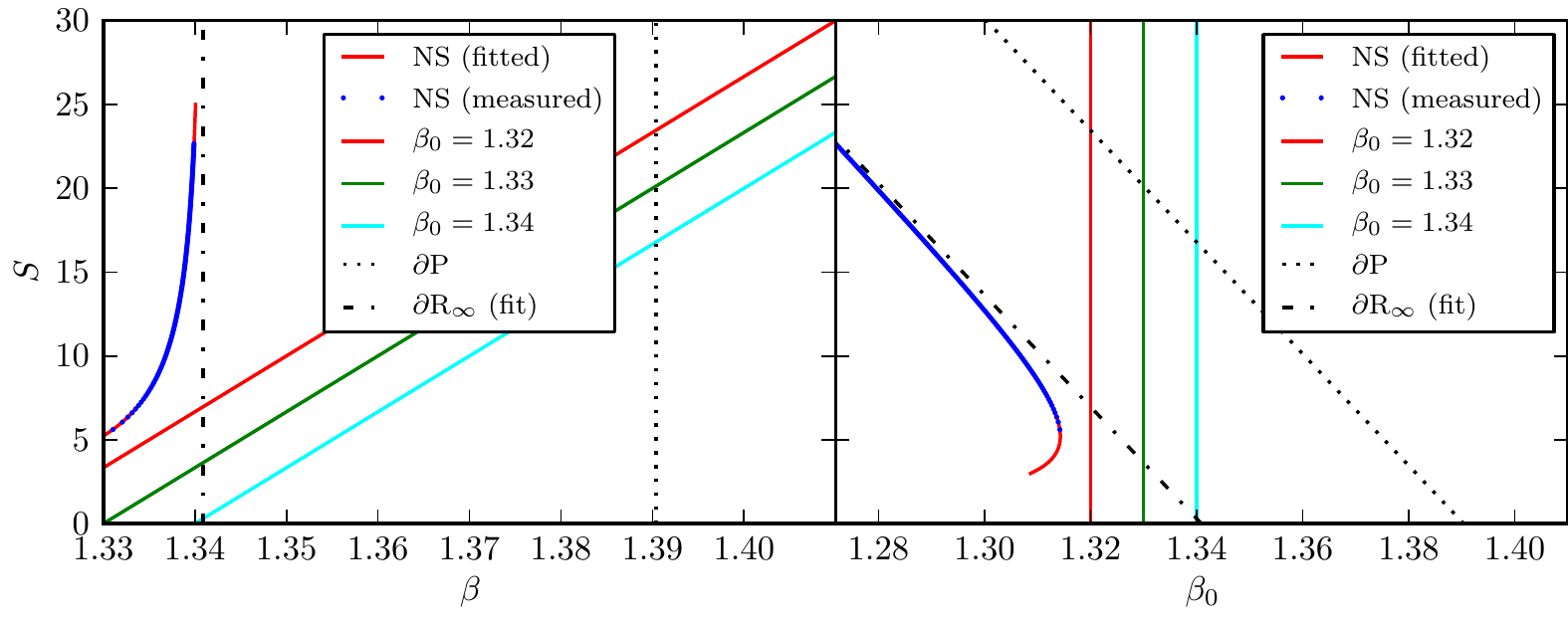}}
  \end{center}
  \caption{\label{fig:control_plane}{\bf (left) The $S$ - $\beta$ plane} with nucleation solution NS , propagation boundary $\partial{\text{P}}$, rotor boundary $\partial_{\textrm{R}_\infty}$ and control lines for the used values of $\beta_0$. {\bf (right) The $S$ - $\beta_0$ plane} with the same quantities.}
\end{figure*}

In Fig.~\ref{fig:control_plane}A, both the rotor boundary $\partial R_{\infty}$
and the propagation boundary $\partial P$ are shown in a bifurcation diagram
for the excitable medium described by Eqs.\,(\ref{eq:fhn1})-(\ref{eq:fhn2}).
We chose $\beta$ as the bifurcation parameter and follow (see Methods) the
branch of the unstable nucleation solution (NS) whose stable manifold separates
the basins of attraction of the homogeneous state and a spiral wave (with two
counter-rotating open ends). The unstable manifold of NS consists of the two
heteroclinic connections, one to the stable homogeneous state and the other to
the traveling wave solution (see Fig.\,\ref{fig:phase_space_sketch}).  The
order parameter on the ordinate in Fig.~\ref{fig:control_plane} is the surface
area $S$ inside the isoclines at $u=0$ of the traveling wave solutions, see
Eq.\,(\ref{eq:S}). We call $S$ the \emph{wave size}.

The mean field control that we introduce by
Eqs.~(\ref{eq:S})-(\ref{eq:control}) establishes a linear feedback  signal of
the wave size $S$ to the threshold $\beta$.  With this linear relation, we
introduce two new parameters, the coupling constant $K$ and $\beta_0$, the
threshold parameter for the  medium without an excited state ($S=0$). Note that
the parameter $\beta_0$ can be also seen as the sum of two threshold values,
the former $\beta$ in Eq.\,(\ref{eq:fhn2}) and an offset coming from the new
control scheme.  While the introduction of the control introduces two new
parameters $\beta_0$ and $K$, at the same time $\beta$ becomes dependent upon
the control, so that we have a total of three parameters.  in this study, we
kept $K$ fixed at $K\!=\!0.003$ and varied $\beta_0$, with a particular focus
on the statistics for $\beta_0\in\left[ 1.32,\! 1.33,\!  1.34 \right]$.

We chose $\beta_0$ as the new bifurcation parameter in the bifurcation diagram
for the full reaction-diffusion model with mean field coupling described by
Eqs.\,(\ref{eq:fhn1})-(\ref{eq:control}), see Fig.~\ref{fig:control_plane}B.
This diagram is a sheared version of the one without mean field coupling in
Fig.~\ref{fig:control_plane}A.  While it is a trivial fact, that the linear
relation in Eq.~(\ref{eq:control}) describes an affine shear of the axises
$(\beta,S)$ of  bifurcation diagram in A to the axises $(\beta_0,S)$ in B,
the fact that the branch of the NS solution can be mapped this way is not.
Firstly, this relies on the way we introduce the feedback term. It just adds a
constant value to the old bifurcation parameter $\beta$, if the solution under
consideration is stationary. Therefore, any stationary solution must exist in
both diagrams being just sheared branches. The same holds true for traveling
wave solutions that are stationary in some appropriate comoving frame
$\xi=x-ct$ with speed $c$. However, not much can be said about the stability of
such solutions, when we introduce the mean field feedback term.

\begin{figure}[b]
\begin{center}
\fbox{\includegraphics[width=\columnwidth]{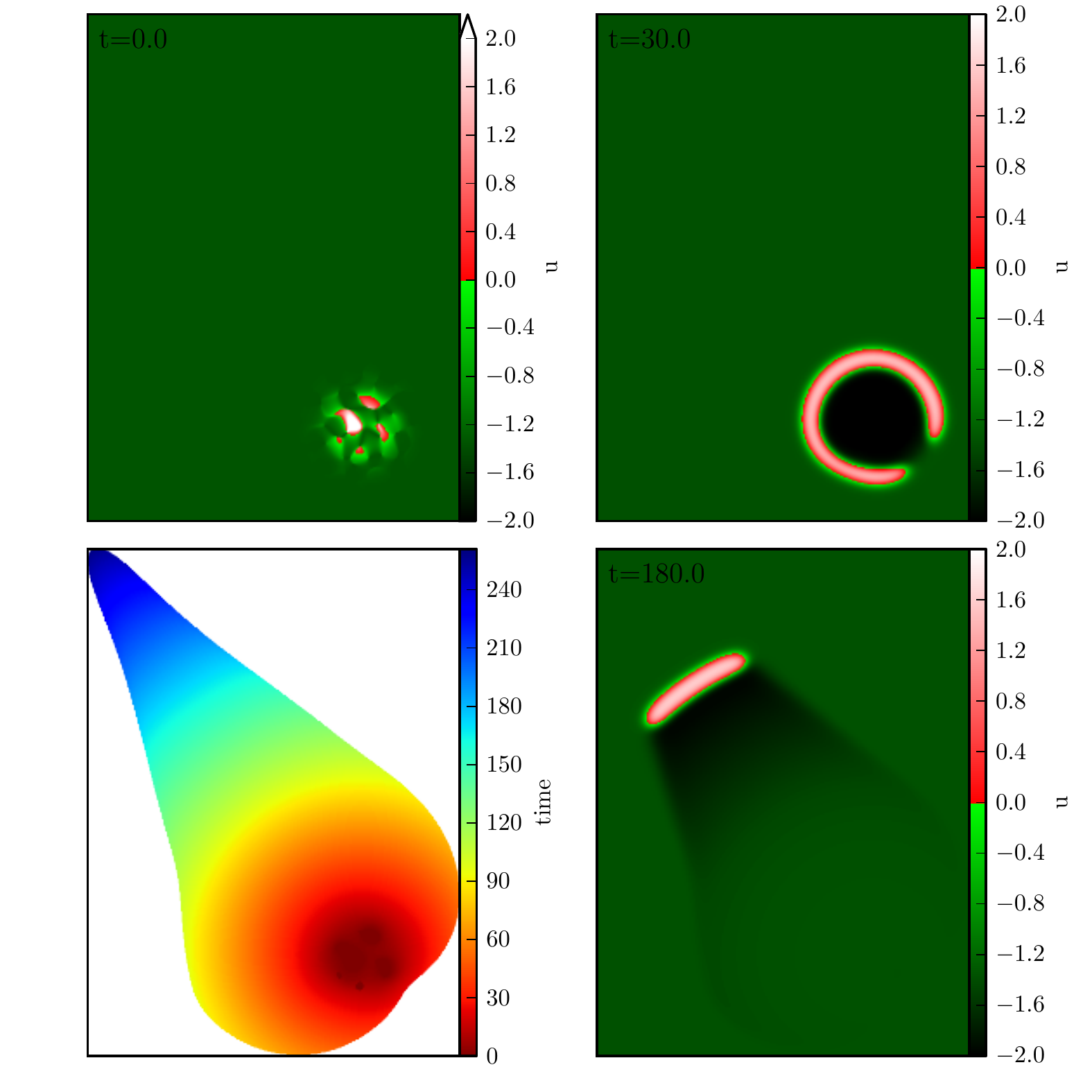}}
\end{center}
\caption{ An example of a transient solution. Initial conditions, i.e.
activator concentration $u$ at $0s$ (upper left), snapshot of
activator-concentration $u$ after $30\,s$ (upper right), after $180\,s$ (lower
right). Time of passing through threshold value $u_0=0$ from below the first
time, i.e. passing of the wave front (lower left).  \label{fig:example_bestia}}
\end{figure}
The branch of the formerly unstable nucleation solution NS
(Fig.~\ref{fig:control_plane}A) folds in Fig.~\ref{fig:control_plane}B such
that two solutions coincide for a given value of $\beta_0$ until they collide
and annihilate each other at a finite value of $S\approx5.5$ for $K\!=\!0.003$.
For the fixed value of $K\!=\!0.003$, the upper branch is a stable traveling
wave solution in the shape of a wave segment, while the lower branch is the
corresponding nucleation solution of this wave segments, as schematically shown
in Fig.~\ref{fig:phase_space_sketch}B. The fact that the upper branch is stable
was confirmed by numerical simulations.   Larger $K$, that is, a less steep
control line in Fig.~\ref{fig:control_plane}A can be seen as a ``harder''
control, because a small given change in $S$ leads to larger variations in the
effective parameter $\beta$. As a consequence, it is difficult to continue by
means of this control the lower part of the branch corresponding to small
traveling wave segments in numerical simulations.

The choice of a parameter regime for this model that shows transient localized
waves and is globally stable with the homogeneous state as the only attractor
is now straightforward. Transient localized waves occur due to a
bottleneck---or ghost behavior---after the saddle-node bifurcation.

\subsection*{Statistical properties}
\label{sec:results}

To examine the typical transient patterns that the system generates, we want to
know how the system responds to arbitrary initial conditions with the
noticeable constraint that the system should initially be in the homogeneous
steady state almost everywhere and the arbitrary perturbations from it are
localized.  As it is not possible to formulate an analytical solution to the
equations for arbitrary initial conditions, the idea is to simulate the
dynamics for (many) different initial conditions.  Ideally, these initial
conditions should be ``equally spaced'' in phase space, in order to obtain
relevant statistics about the different evolution possibilities.  The problem
that arises at this point is that an initial condition of this system is not
only living in an infinite dimensional space (an initial condition would be
given by two $C^2$-functions $\mathbb{R}^2 \rightarrow \mathbb{R}$) but because
of the nonlinearity of the equations, the set of all solutions is not even a
vector space. To our knowledge, there is no helpful mathematical structure that
could guide us in choosing our initial conditions.  To attack this problem, we
take a set of patterns, which are parameterized by a finite number of
parameters and scan through these parameter. For details on the patterns see
Methods.

Of course, in characterizing the solutions, the same problem arises and
appropriate characteristic parameters for the solutions have to be defined.

\begin{figure}
\begin{center}
\fbox{\includegraphics[width=\columnwidth]{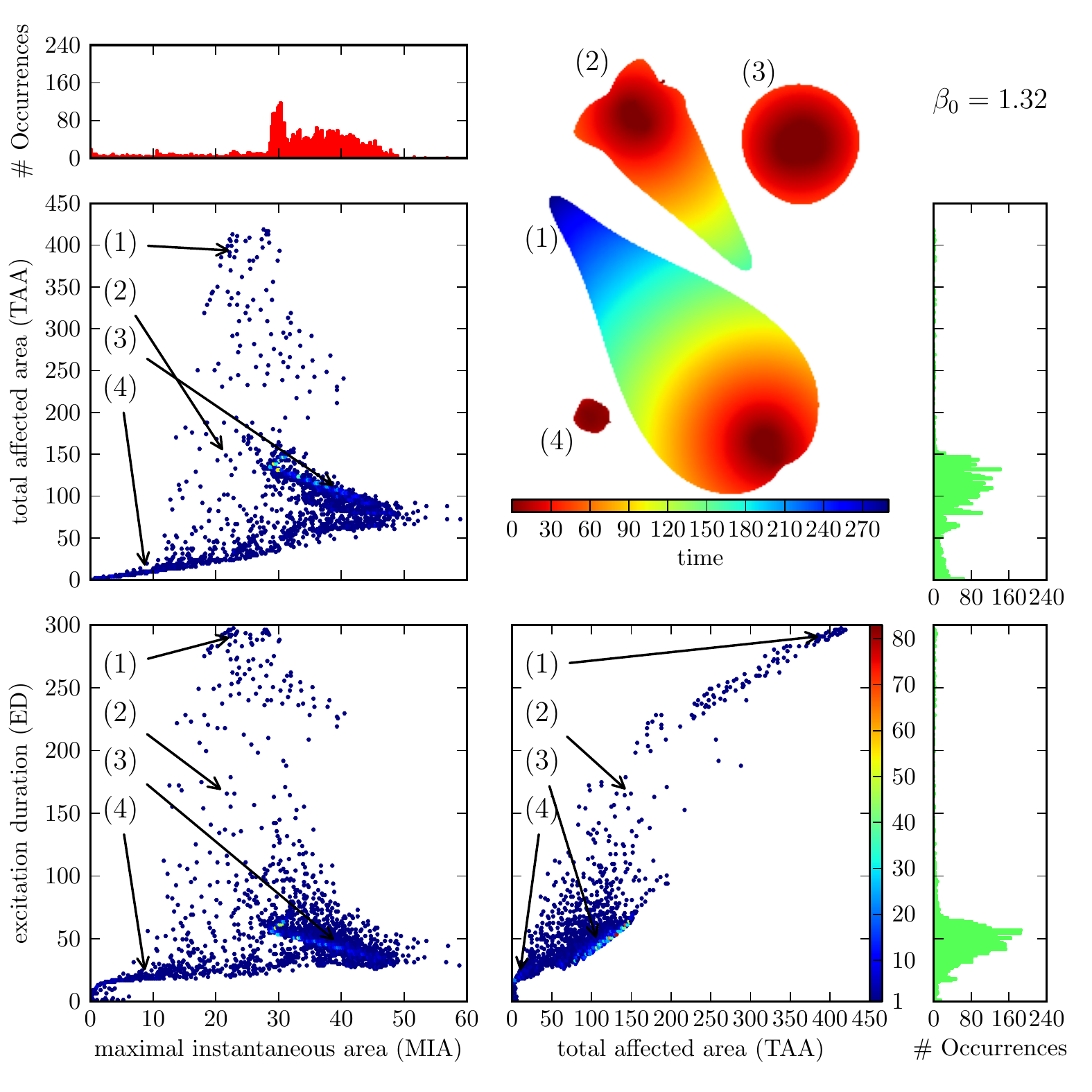}}
\end{center}
\caption{
{\bf Distribution of solutions}  for the control close to $\partial {}_\text{R}$. 
\label{fig:distr_near}}
\end{figure}

\begin{figure}
\begin{center}
\fbox{\includegraphics[width=\columnwidth]{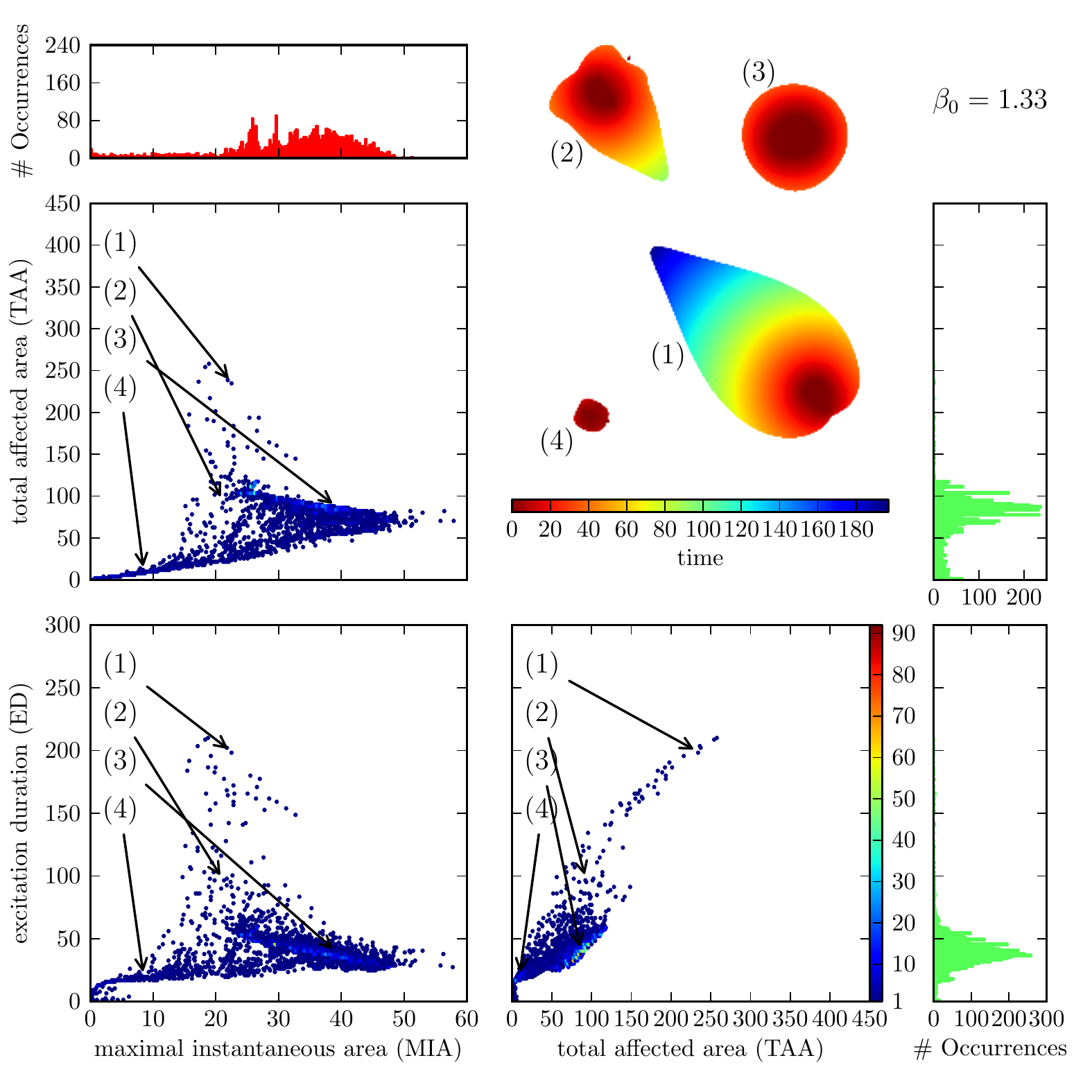}}
\end{center}
\caption{
{\bf Distribution of solutions}  for the control line at an intermediate distance
to $\partial {}_\text{R}$. \label{fig:distr_middle}}
\end{figure}

\begin{figure}
\begin{center}
\fbox{\includegraphics[width=\columnwidth]{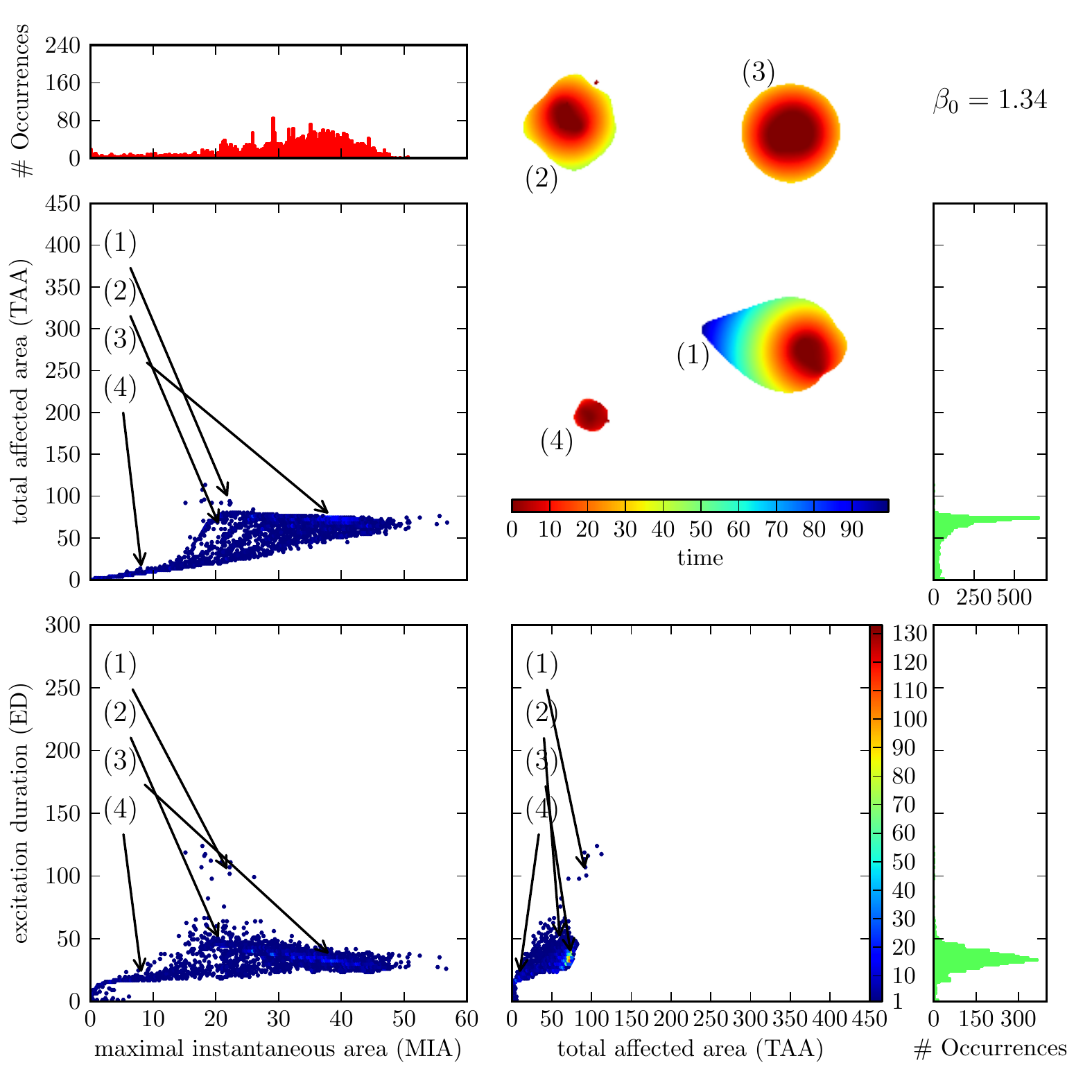}}
\end{center}
\caption{
{\bf Distribution of solutions}  for the control far away from $\partial {}_\text{R}$. 
\label{fig:distr_far}}
\end{figure}

\begin{figure*}
\begin{center}
\fbox{\includegraphics[width=\textwidth]{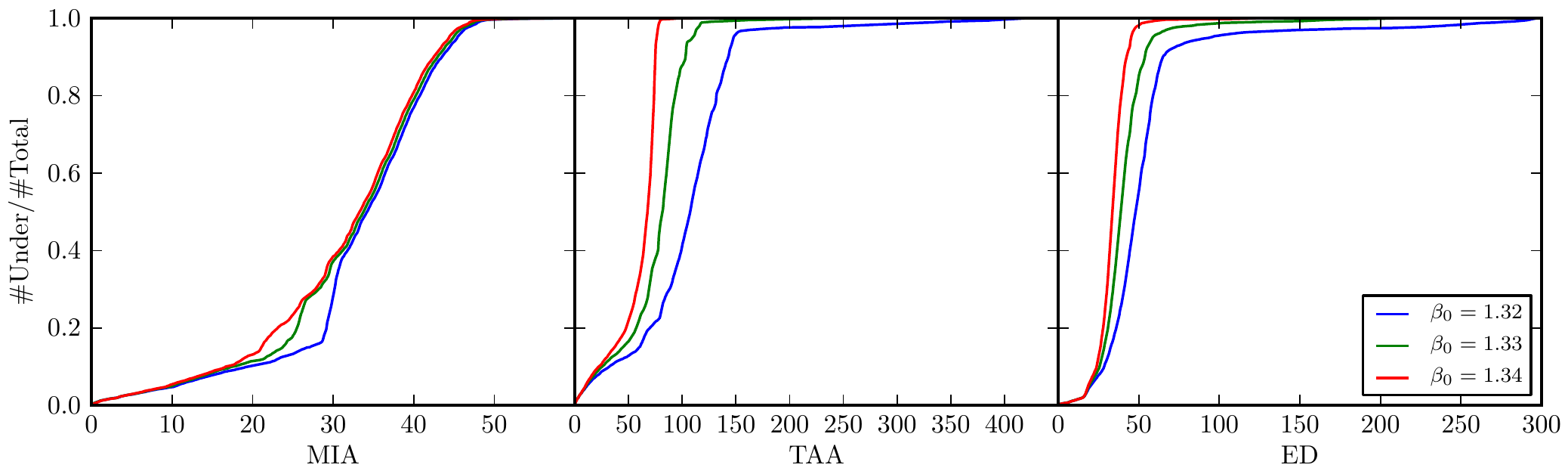}}
\end{center}
\caption{
{\bf Cumulative distribution functions} for the different classification 
parameters.\label{fig:cum_distr}}
\end{figure*}

To explain the three parameters we have chosen for the solutions and why they
suit this problem, it is helpful to have a look at the lower left part of
Fig.~\ref{fig:example_bestia}, in which an example solution is displayed.  The
first parameter we choose is the maximal area in which such a solution has
activator concentration over a certain threshold level at one instant of time,
termed maximal instantaneous area (MIA). The threshold level is taken to be
$u=0$, although this is the same threshold as used to define $S$, this is
rather convenience than necessity.  The second parameter is the total area that
has experienced an activator concentration above this level at some time during
the course of the solution, termed total affected area (TAA).  The third
parameter is the time, during which the area of activator concentration above
threshold is non-zero, termed the excitation duration (ED).  Of course, the
exact value of all these parameters for one single solution depends on the
choice of threshold. For once, the threshold value has to be chosen such that
after the activator concentration has fallen below it, no secondary excitation
will be generated.

The example solution depicted in Fig.~\ref{fig:example_bestia} is a
comparatively long lived solution. It starts out very symmetrically
(circularly) shaped, at one instant of time it breaks open into a discontinuous
wave and a shape of the front develops, which is similar to that of a
particle-like wave but because of the chosen control parameters, it shrinks in
time and vanishes in the end. Because at the point when the circle breaks open,
a comparatively large area is affected, it takes some time until it vanishes
and the resulting TAA is relatively large. So this example solution has large
ED, large MIA and large TAA. If the circle had not broken open at all, the
control would have made the threshold value very large and the solution would
have collapsed very quickly because of the propagation boundary $\partial P$,
such that the ED and the TAA would have been short, whereas the MIA would have
been large.  Other prototypical courses of solutions take place for instant,
when the initial conditions affect the activator over a larger area but only in
the middle of the area, the value is high enough to start a solution. In the
surrounding area, the activator level is not high enough for that but the
increased activator concentration leads to a rise in inhibitor concentration
until the time the front reaches those parts and as a consequence, the solution
vanishes early, having small ED, small TAA and small MIA as a consequence.

We did the simulation for three different adjustments of the control force,
successively going farther and farther away from the bifurcation point. Each of
these simulations were started using ~8000 initial conditions generated in a
manner that is described in Methods. Each of these initial conditions resulted
in a solution that was classified according to the three parameters mentioned
above. The solutions that did not result in any excitation at all
(ED=MIA=TAA=0) were discarded. The density plots according to the
classification parameters are shown in
Figs.~\ref{fig:distr_near},~\ref{fig:distr_middle},~\ref{fig:distr_far}.  First
of all, though the distribution of the solutions varies significantly, the
number of solutions that represent an excitation hardly varies at all (4171 for
small, 4183 for intermediate, 4182 for large distance from the bifurcation
point), the symmetric difference between the sets of initial conditions that
lead to an excitation contains between 5 and 19 solutions. From this we can
also deduce that the set of initial conditions that lead to an excitation does
not significantly depend on the choice of control parameters.


When looking at Fig.~\ref{fig:distr_near} one notices a clustering of the
solutions in certain regions of the classification parameters. In the section
that depicts the TAA against the MIA, we notice three coarse clusters. Cluster
I, the largest with high MIA and comparatively low TAA; cluster II, one that is
less populated with low MIA and low TAA; and cluster III, one that is very
sparsely populated with intermediate MIA and high TAA.  The boundaries between
these clusters are not very sharp. One could think that a solution that affects
an overall large area (high TAA) will also affect a large area at one instant
of time (high MIA).  From looking at the mentioned clusters, one sees that this
is not the case, the solutions with the highest MIA have all comparatively low
TAA (clusters I and II) and the ones that have a high TAA only achieve an
intermediate MIA (culster III).


A partial explanation for this can be read off from the depiction of TAA
against ED. All solutions that have a large TAA are also solutions that have a
large ED, i.e., cluster III is distinct also in this plane. More than that, the
dependence seems to be almost linearly. This is reminiscent of the localized
particle-like wave solutions. For these, the area that is affected grows
linearly in time because the area that these solutions occupy at one instant of
time is constant.


The two clusters I and II that we observed merge to one in this plane of
projection because they differ only very little in ED.  This can also be noted,
when comparing the planes MIA vs.\ TAA and MIA vs.\ ED, also here the cluster
III with high TAA translates to a cluster with high ED and the cluster I with
high MIA and comparatively low TAA moves closer to cluster II with both low ED
and MIA.


When varying the $\beta_0$ parameter of the control force, the distribution of
solutions in MIA-TAA-ED-space changes drastically.  Upon raising the $\beta_0$
parameter from $\beta_0= 1.32$ over $\beta_0=1.33$ to $\beta_0=1.34$, the
system is put more and more into the subexcitable regime and the solutions are
less and less affected by the ghost behavior (saddle-node bifurcation), see
Fig.~\ref{fig:control_plane}.  This is noticeable by observing that the cluster
with high TAA / high ED becomes less pronounced and vanishes almost completely
for $\beta_0=1.34$. This can be understood as an interplay between the mean
value of MIA in cluster III at about 25 and $S$ at the propagation boundary (at
$\partial P$, $S\approx24$, $S\approx20.75$, and $S\approx17.5$ for $\beta_0=
1.32$, $\beta_0=1.33$, and $\beta_0=1.34$, respectively). For the control line
farthest away from the saddle-node bifurcation ($\beta_0=1.34$), $\partial P$
is below even the smallest values of MIA  in cluster III. Note that the value
of $S$ at the ghost is about 6, well below the propagation boundary.  Also the
other two clusters merge though there still exist solutions with high and with
low MIA, but the transition is much more fuzzy than it was before.

In Figs.~\ref{fig:distr_near},~\ref{fig:distr_middle}~and~\ref{fig:distr_far},
we have included a little `bestiary' to illustrate the typical courses of
solutions in the respective clusters and their change upon varying the
parameter $\beta_0$, the initial conditions for solutions 1-4 in these figures
are always the same. From this arbitrarily chosen selection, we see that the
MIA of each solution hardly changes between the $\beta_0$ values.  Whereas the
change of TAA and ED always go hand in hand and---depending on the
cluster---can be up to fourfold for the chosen range of $\beta_0$.

One could argue that the formation of clusters is an artefact of the choice of
initial conditions. There is no simple answer to this. As mentioned, it is not
possible to examine the complete set of initial conditions. Neither does this
set carry a helpful structure which would allow a sensible `equidistant'
sampling. This is the reason why we made the mentioned choice of initial
conditions.  For testing purposes, we also tried different schemes for the
generation of initial conditions and found the same distribution of clusters
qualitatively.

In Fig.~\ref{fig:cum_distr}, we have plotted the cumulative distribution
functions for the three classification parameters and the three choices of mean
field control. From this picture we see, that the distribution of the MIA is
only hardly influenced by the choice of control. This is very different for TAA
and ED. For the TAA for example there are values (around 75), where for one
choice of control the majority of solutions is below and for another choice the
majority is above. For example, the fraction of values below TAA=80 is 0.995
for $\beta_0=1.34$ and 0.216 for $\beta_0=1.32$. Also, we see that the
cumulative distribution function for the TAA converges to 1 much slower, the
closer the control is to the saddle-node bifurcation. This means, that more
solutions with high TAA exist for these choices of control.

\section*{Discussion}
\label{sec:discussion}

In this section we discuss three subjects related to the intended application
to migraine pathophysiology. Firstly, the possible congruence between the
prevalence of migraine subforms with the statistical properties of the wave
patterns we observed; secondly, the possible physiological origin of the
inhibitory feedback control; and thirdly, novel therapeutic approaches.  We
start with a discussion of the approach of using a canonical model.

\subsection*{Canonical model and generic parameters for weakly excitable media}

More realistic models of SD are given by a conductance-ion-based models of
SD\cite{KAG00,SHA01,MIU07} with up to 29 dynamics variables. The fast
inhibitory feedback that we suggest can be modeled in addition by neural field
models\cite{BRE12}. Such large-scale models of brain structure, including
lateral connectivity, are available but still require enormous computer
capabilities. We will argue here, in which sense our model is canonical for the
problem we attack.

Generally speaking, an excitable medium is a spatially extended system with a
stable homogeneous steady state being the quiescent state and one or many
excited states that develop after a sufficient perturbation from the quiescent
state (Fig.~\ref{fig:phase_space_sketch}A).  The excited states are traveling
wave solutions that propagate with a stable profile of permanent shape
(possibly with some temporal modulation, such as breathing or meandering). To
study generic features of  an excitable medium, the simulations are often
carried out in the reaction-diffusion system given by
Eqs.\,(\ref{eq:fhn1})-(\ref{eq:fhn2}), the popular FitzHugh-Nagumo kinetics.
Originally, the FitzHugh-Nagumo kinetics were a caricature of the
electrophysiological properties of excitable membranes \cite{FIT61,NAG62}, but
these equations with $D=0$ became a canonical model of {\em local} excitability
of type II (based on Hopf bifurcation, either supercritical with subsequent
extremely fast transition to a large amplitude limit cycle, named canard
explosion\cite{WEC05}, or subcritical \cite{ERM98}), and also, for $D\neq0$, of
{\em spatial} excitability \cite{WIN91}, sometimes including diffusion in the
second inhibitory species, which we do not consider here. Because we are
considering transient behavior originating from a high threshold regime
(towards weak excitability), the classification of local excitability in type I
and II (based on the transition at vanishing threshold, i.e., into the
oscillatory regime) is not relevant. Furthermore, it is not clear whether this
classification carries over in a meaningful way to the dynamics of spatially
extended systems. 

We consider the set of Eqs.\,(\ref{eq:fhn1})-(\ref{eq:fhn2}) as canonical for
two reasons. First, because the activator Eq.\,(\ref{eq:fhn1}) has the simplest
polynomial form of bistability. Note that Eq.\,(\ref{eq:fhn1}) was for this
reason originally suggested by Hodgkin and Huxley as the first mathematical
model of the potassium dynamics in SD. It was published by Grafstein, who also
provided experimental data supporting such a simple reaction-diffusion scheme
for the front dynamics \cite{GRA63}. Second, the inhibitor Eq.\,(\ref{eq:fhn2})
has a linear rate function, in fact, the rate function is only of a function of
the activator $u$. This is the simplest inhibitor dynamics needed for pulse
propagation. By neglecting an additional linear term $-\gamma v$ in the
inhibitor rate function, we limit the origin of excitability to the case of a
supercritical Hopf bifurcation with subsequent canard explosion and avoid the
bistable regime that exits in the subcritical case. The subcritical Hopf
bifurcation occurs only in a narrow regime when $\gamma$ is close to 1 and $\beta$ close to 0. We have
tested some simulations with $\gamma=0.5$ with similar results. 

As a consequence of our assumptions about the model being in this canonical
form, only two parameters exits, $\beta$ which is associated with the threshold
and the $\varepsilon$, the time scale separation of activator and inhibitor
dynamics. Of course, the choice of parameters can be quite different, a common
choice is $\alpha$ in the cubic rate function $f(u)=u(u-\alpha)(u-\alpha)$ but
there are only two free parameters or two equivalent groups of parameters. So
there are the same bifurcations in the parameter planes $(\varepsilon,\beta)$
or $(\varepsilon,\alpha)$, but to map the dynamics between equivalent groups of
parameters might involve changes in time, space and concentrations scales.  

In particular the question of how the incidence of MA is reflected in the
distance to the saddle-node bifurcation, involves a measure on the parameter
space, which we have suggested to get from pharmacokinetic-pharmacodynamic
models\cite{DAH07a}. 

\subsection*{Application to migraine pathophysiology}

The cause of the neurological symptoms in migraine with aura (MA) is the
phenomenon of cortical spreading depression (SD)\cite{LEA45,OLE81,LAU87,HAD01}.
Whether SD is also a key to the subsequent headache phase is an open question,
in particular, in cases of migraine without aura (MO). If SD occurs in MO, it
must remain clinically silent \cite{AYA10,EIK10} or---by definition of
diagnostic criteria---neurological symptoms must last less than 5min. Of
course, the transient nature of SD poses challenging problems in clinical
observability, in particular for objective measures by means of non-invasive
imaging when clinical symptoms do not even indicate the aura phase with SD.
The aura is usually, though not always, before the headache phase.  Attacks
observed with non-invasive imaging are usually triggered, which also could
cause a trigger-specific bias. One well-documented case of a spontaneous
migraine headache supports the contested notion of 'silent aura', because
blood-flow changes where observed that were most likely the result of
SD\cite{WOO94}.

We suggest a qualitative congruence between the prevalence of  MO and MA with
the statistical properties we found in the transient response properties.  We
do not suggest that all MO attacks are related to SD nor that pain formation in
MA is exclusively caused by SD. Rather that SD is one pathway of pain formation
in MO and MA. We refer to this pathway as the ''spreading depression''-theory
of migraine\cite{LAU87}.  The ''migraine generator''-theory (MG), a dysfunction
in a central pattern generator in the brainstem that modulates the perception
of pain, is for various reasons not less plausible\cite{AKE11}. Some of the
seemingly conflicting and controversially discussed evidence is probably
resolved when one considered the basis of the classification of migraine. We
currently have a symptom-based classification for migraine with possibly
overlapping etiologies for individual subforms. In the light of an
etiology-based classification with possibly overlapping symptoms the conflicts
seem less puzzling to us.  We also need to investigate an interplay of SD and
MG, namely to which degree MG modulates pain traffic from SD generated in the
intracranial tissues.

\begin{figure}
\begin{center}
\fbox{\includegraphics[width=\columnwidth]{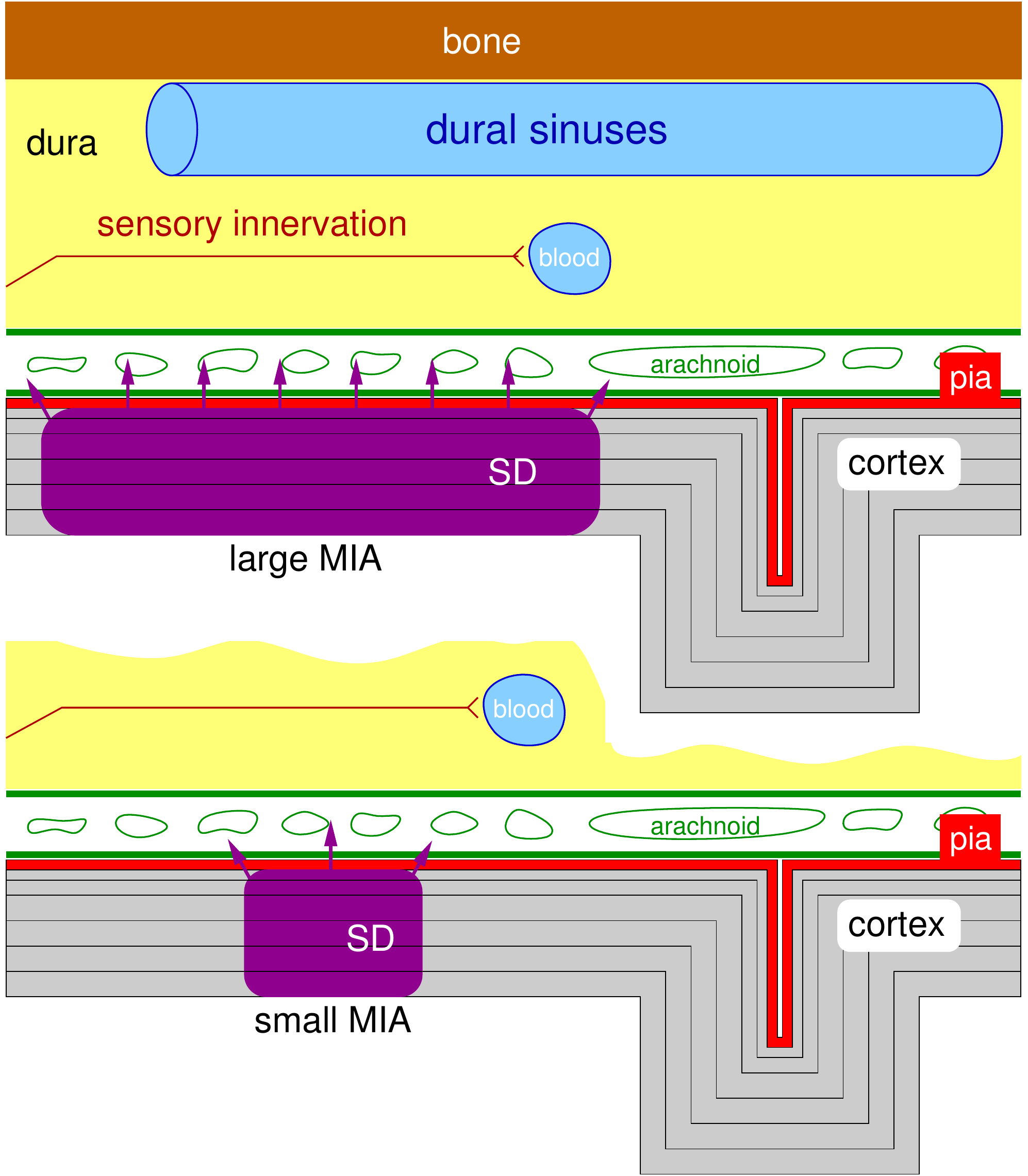}}
\end{center}
\caption{{\bf Schematic representation of cross section of cortex, meninges and skull.}
The leptomeninges refer to the pia mater and arachnoid membrane.  SD releases
noxious substances with increased blood flow thought to diffuse outward.
Activation of pain pathways can depend on MIA. \label{fig:cortex_meninges_mia}}
\end{figure}

The cortex is not pain sensitive.  There are detailed investigations how SD in
the cortex  can cause pain via pain sensitive intracranial tissues and
subsequent activation in the trigeminal nucleus caudalis in the
brainstem\cite{MOS93a,BOL02}, but cf. \cite{ING97,MOS98}.   The qualitative
congruence between the prevalence of  MO and MA with the statistical properties
we found in the transient response properties is based on the following
assumption on the geometrical layout: In the initial phase of cortical SD, with
increased blood flow (hyperemic phase), a local release of noxious substances
(ATP, glutamate, K$^+$, H$^+$) are thought to diffuse outward in the direction
perpendicular to the cortex into ``the leptomeninges resulting in activation of
pial nociceptors, local neurogenic inflammation and the persistent activation
of dural nociceptors which triggers the migraine headache''\cite{ZHA10}, but
for issues concerning the blood brain barrier system cf. \cite{TFE11}. If
diffusion vertical to the affected area is critical, size and shape of this
area should play a critical role, see Fig.~\ref{fig:cortex_meninges_mia}. 
This suggests that SD waves activate nociceptive mechanisms dependent upon a
sufficiently large instantaneously affected cortical area, i.e., large MIA.

The aura phase on the other hand must clearly correlate with
long and large enough cortical tissue being affected to notice the neurological
deficits. In particular, because the very noticeable visual symptoms often
start where the cortical magnification factor  is large, so that only if they
move into regions of lower magnification they get magnified by the reversed
topographic mapping \cite{DAH03a}. The seemingly contested notion of MO
(migraine without aura) with silent aura is also resolved.

\begin{figure}
\begin{center}
\fbox{\includegraphics[width=\columnwidth]{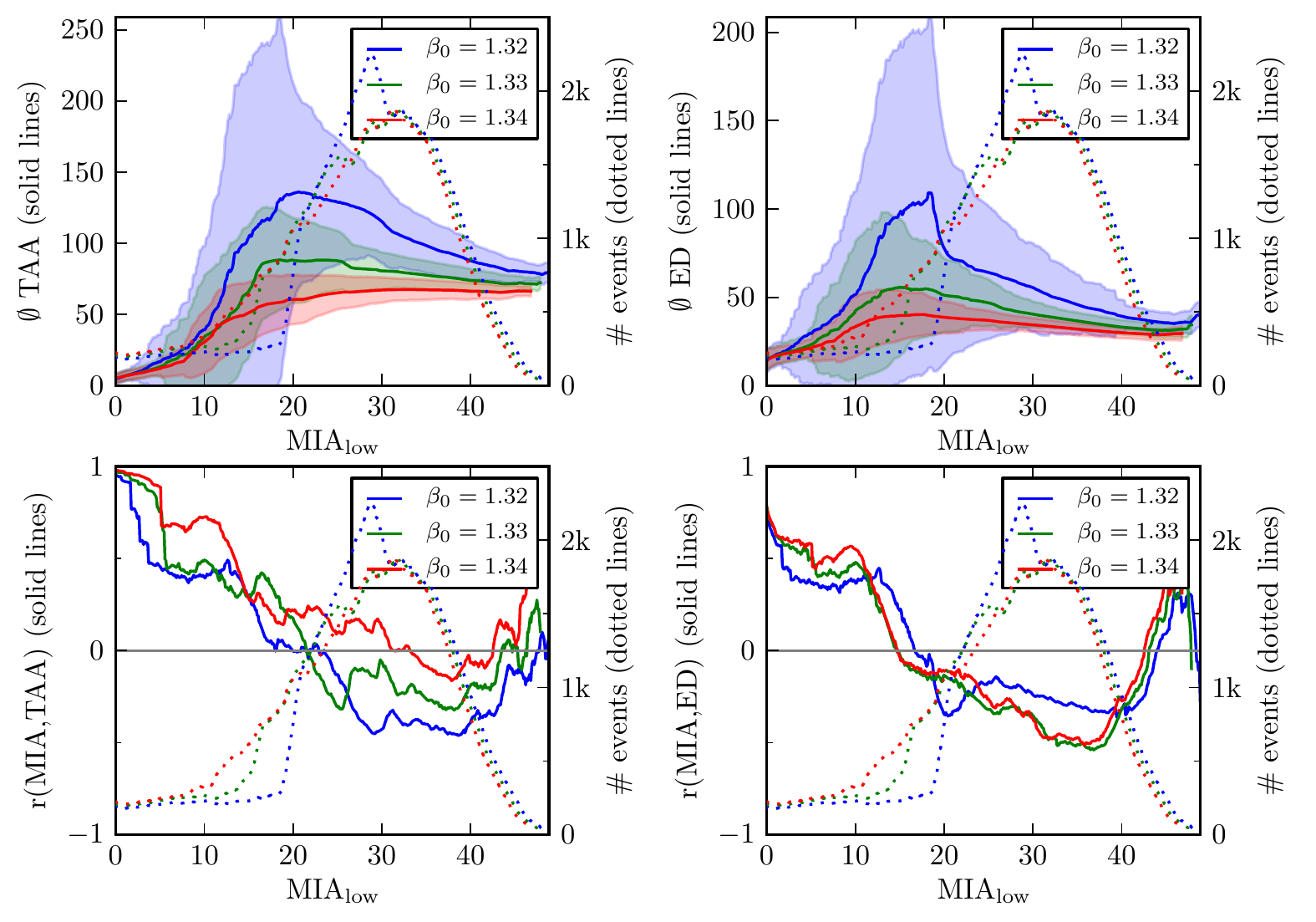}}
\end{center}
\caption{{\bf Statistical analysis of output data.} 
  For all four pictures we took all data points with MIA in the interval [MIA${}_\text{low}$, MIA${}_\text{low}$+10] (``sliding window'') and analyzed the connection with TAA {\bf (left column)} and ED {\bf(right column)}. In the {\bf(upper row)}, the average value is plotted with \emph{solid lines}, the area of one standard deviation around this value is shaded. In the {\bf(lower row)}, the correlation coefficient between MIA and the respective quantity is plotted.
In all plots, the \emph{dotted lines} indicates the number of events for the respective interval with MIA${}_\text{low}$.
 \label{fig:statistical_evaluation}}
\end{figure}

A view at the connection between MIA and TAA as well as MIA and ED in our model is shown in Fig.~\ref{fig:statistical_evaluation}.
It shows that in the range of high MIA the average values for TAA and ED are becoming smaller. From Fig.~\ref{fig:statistical_evaluation} we can also read off that the range with the most events is in the regime of relatively high MIA (around 30) and significantly after the peak of ED resp.\ TAA. Moreover in the range with most events, the correlation coefficient r(MIA, ED) is always negative and the correlation coefficient r(MIA,TAA) is mostly negative. All these effects are stronger, the closer the control line is located to the saddle-node bifurcation.

From these statistical correlations between MIA and TAA resp.\ ED and the distribution of the number of events, one could speculate that cases of MA are more rare and the quality of the headache in these cases might be less severe.
This is exactly what has been reported in the medical literature\cite{RAS92}.

While the number of events with high ED and high TAA is influenced by the distance of the control line to the saddle node bifurcation, the number of events with high MIA is much lesser affected.
So in a way, the distance to the saddle node bifurcation controls the prevalence of MA in our model, while the prevalence of MO is not much affected.

\subsection*{Inhibitory feedback and neurovascular coupling}

This naturally rises the question of the physiological origin of the inhibitory
feedback control. The hyperemic phase engulfs large regions of the human
cortex\cite{OLE81}, while we suggest in this study the homeostatic breakdown
directly due to SD is much more limited in extent. A fast spreading increased
neural activation in adjacent cortical areas could represent synaptic
activation through feed-forward and feedback circuitry. This was suggested by
Wilkinson \cite{WIL04}. This would in turn extend the area of the hyperemic
phase towards tissue that is not yet recruited into the SD state and this
mechanism has therefore a neuroprotective effect by an increased blood flow
which we mimic by the inhibitory mean field feedback.

The coupling between neural activity and subsequent changes in
cerebral blood flow, called neurovascular coupling, has a significant time
delay in the order of seconds, which we ignore for the sake of simplicity in our model.

\subsection*{Model-based control by neuromodulation}

We briefly discuss model-based control and means by which neuromodulation
techniques may affect pathways of pain formation and the aura phase.

The emerging transient patterns and their classification according to size and
duration offer a model-based analysis of phase-dependent stimulation protocols
for non-invasive neuromodulation devices, e.g.\ utilizing transcranial magnetic
stimulation (TMS)\cite{LIP10}, to intelligently target migraine.  For instance,
noise is a very effective method to drive the system back into the homogeneous
steady state more quickly.  In general, responses of nonlinear systems to noise applied when
the system is just before or past a saddle-node bifuraction are well studied.
Before the saddle-node on limit cycle bifuraction, the phenomenon of coherence
resonance (CR) describes that a certain amount of noise makes responses most
coherent\cite{LIN04}.  Behind the saddle-node bifurcation on a limit cycle the time the flows
spend in the bottleneck region of the ghost is shortened\cite{STR94a}.  However,
noise would, according to our model, mainly positively affect ED and TAA, that
is, the aura, while it could even worsen the headache, if applied early during
the nucleation and growth process. Therefore, TMS using noise stimulation
protocols, which is currently considered, should be applied only some time
after first noticing aura symptoms.

Headaches are not generally considered appropriate for invasive neurosurgical
therapy, but when all else fails---preventives, abortives, and pain
management---invasive brain stimulation techniques are also considered, e.g.
occipital nerve stimulation (ONS)\cite{SIL12,DIE12}. So model-based control
will become increasingly important. Also the importance of modeling related
epileptic seizure dynamics as spatio-temporal transient patterns, has been
suggested in a recent paper\cite{BAI12}. Model-based control of Parkinson's
disease, is already considered, yet Schiff remarks quite
correctly\cite{SCH10h}: ``It seems incredible that the tremendous body of skill
and knowledge of model-based control engineering has had so little impact on
modern medicine. The timing is now propitious to propose fusing control theory
with neural stimulation for the treatment of dynamical brain disease.''

We suggest to consider migraine as a dynamical disease that could benefit from
model-based control therapies.

\section*{Methods} 
\label{sec:methods}

As a generic model for our excitable medium we use the well known FitzHugh-Nagumo equations \cite{IZH06}, augmented by a diffusion term for the activator variable:
\begin{align}
\varepsilon \frac{\partial u}{\partial t} 
                     &=  u -\frac{1}{3} u^3 -v + \nabla^2u   \label{eq:fhn1}      \\
\frac{\partial v}{\partial t}
                     &=  u+ \beta.        \label{eq:fhn2} 
\end{align}
The parameter $\varepsilon$ separates the timescales of the dynamics of the activator $u$ and the inhibitor $v$. $\varepsilon$ is taken to be small. In the present work, we use a value of $\varepsilon=0.04$. 
The parameter $\beta$ is a threshold value which determines from which activator level on the inhibitor concentration is rising. The local dynamics of \eqref{eq:fhn2} (i.e. without the diffusion term) is oscillatory for $\left|\beta \right|<1$ and excitable for  $\left|\beta \right|>1$. At  $\left|\beta \right|=1$ the local dynamics undergo a supercritical Hopf-bifurcation. We choose a value of $\beta=1.1$ throughout this work.
To integrate \eqref{eq:fhn2}, we used a simulation based on spectral methods \cite{CRA06} and adaptive timestepping.

We define the (instantaneous) wave size as the area with activator level $u$ over a certain threshold $u_0$:
\begin{equation}
S(t)         :=   \int\!\!\!\int \Theta\left(u(x,y,t)-u_\text{threshold}\right)\,\mbox{d}x\mbox{d}y \label{eq:S},
\end{equation}
where $\Theta$ is the Heaviside function  and $u_\text{threshold}=0$.

Equations \eqref{eq:fhn2} are a paradigmatic model of an excitable medium \cite{MIK90}. It possesses a stable homogeneous solution as well as stable excited states (pulses, spirals or double spirals) cf.~\cite{DAH08,DAH09a} The boundary separating the basins of attraction of these types of solution consists of unstable so-called `nucleation-solutions', which are areas of excitation which are traveling at uniform speed. The size in the sense of \eqref{eq:S} of these solutions is plotted against the parameter $\beta$ in Fig.~\ref{fig:control_plane}. To measure this line $\partial_\text{R}$, called the `rotor boundary', we used a pseudo-continuation procedure. Which is described below.

Making the parameter $\beta$ dependent on the wave size $S$ adds a mean field control to the system.
\begin{equation}
\beta                = \beta(t) =  \beta_0  + K \cdot S(t)  \label{eq:control},
\end{equation}

where $K$, $S_0$ and $\beta_0$ are control parameters.

If the control line defined by \eqref{eq:control} intersects $\partial_\text{R}$, the point of intersection with higher $S$ is stabilized, cf.~\cite{KRI94}. 

The aim of the present work is to shed light on the transient behavior, occurring when the control line~\eqref{eq:control} is close to $\partial_\text{R}$ but does not intersect it. 

To account for the imprecision in the rotor boundary of the simulation and the exact rotor boundary $\partial {}_\text{R}$, we measured the rotor boundary in our simulation using a pseudo-continuation procedure. For this, we set the control such that it intersects  $\partial {}_\text{R}$ and thus stabilize an otherwise unstable solution on it. Letting the simulation run until the system has stopped fluctuating, saving the ($\beta$,$S$)-pair, changing the control slightly and doing things over yields  points of $\partial {}_\text{R}$ in our system.
From this measured $\partial {}_\text{R}$ and the propagation boundary, inferred from continuation in 1D, we chose 3 suitable control lines which were used for simulations in this work:
\begin{align}
  K        &= 0.003     \nonumber \\
  \beta_0  &\in \left[ 1.32,\! 1.33,\! 1.34 \right] \label{eq:control_pars}
\end{align}

For the purpose of visualization of the saddle-node bifurcation occurring in the bifurcation diagram of the system with mean field control (right of Fig.~\ref{fig:control_plane}, we fitted the measured branch of nucleation solutions to a function of the form $\beta  = a + \frac {b}{c S + S^2}$ which also allows us to obtain an approximation for the $\beta$ value of the rotor boundary $\partial{}\textrm{R}_\infty$.

As mentioned in section~\ref{sec:results}, we need an appropriate sampling of
initial conditions for~\eqref{eq:fhn2}, ideally being equidistantly sampled in
some distribution. As was also mentioned, the set of all initial conditions for
this system does not---to our knowledge---carry a helpful mathematical
structure which allows us to achieve this aim easily.  In order to attack this
problem, we turned to the physiological origin for that we chose this model.

\begin{figure}
\begin{center}
\fbox{\includegraphics[width=\columnwidth]{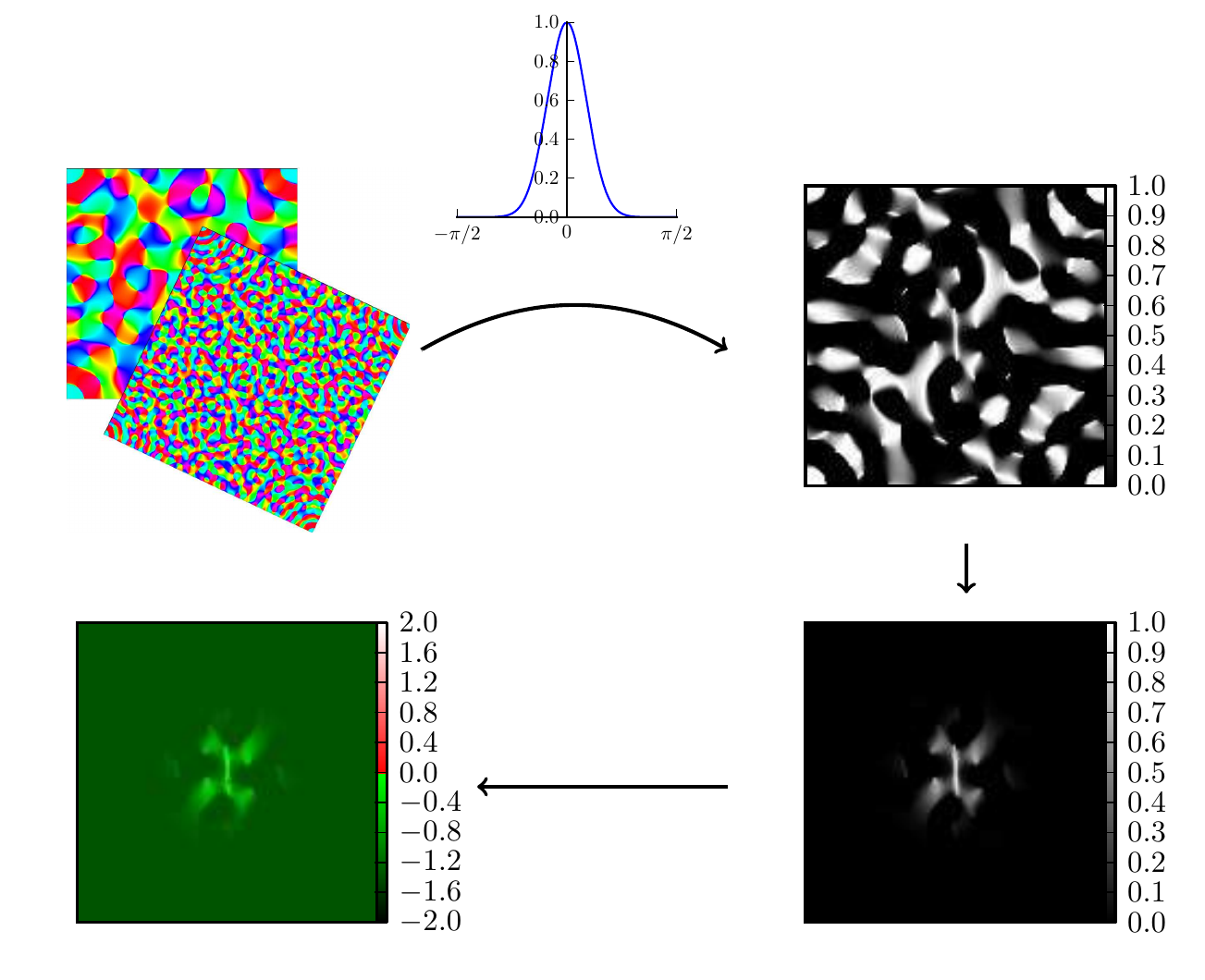}}
\caption{To construct initial conditions from artificially generated pinwheel maps, we first took such a pinwheel map with a certain \emph{scaling} (upper left), then we chose a selection of excited orientations by means of a gaussian. The width of the gaussian gives the selection \emph{depth} (upper right). After that, we masked the result spatially with another gaussian distribution that is radially symmetric (lower right). The width of this gaussian gives the third parameter, the \emph{size} of the pattern. Finally the result is scaled, giving rise to the fourth parameter, we called the \emph{excess} and added to the activator variable in the homogeneous state (lower left). The inhibitor variable is put into the homogeneous state. \label{fig:ini_con_schematic}}
\end{center}
\end{figure}
A set of initial conditions should naturally reflect plausible spatial
perturbations of the homogeneous steady state of the cortex. This can be
achieved by defining localized but spatially structured activity states on
large-scales, i.e., on the order of millimeters. Such pattern are obtained from
cortical features maps (see Fig.~\ref{fig:ini_con_schematic}) by sampling three
parameters (\emph{scaling}, \emph{depth}, and \emph{size}) that define patches
of lateral coupling in theses maps. A fourth parameter (\emph{excess})
determines the amplitude of the perturbation. In the following, we first describe 
the rational behind using a cortical feature map and then the sampling.

We focus on a cortical feature map in the primary visual cortex (V1) called
pinwheel map.  V1 is located at the occipital pole of the cerebral cortex and
is the first region to process visual information from the eyes. Migraine aura
symptoms often start there or nearby where similar feature maps exist.

In V1, neurons within vertical columns (through the cortical layers) represent  by
their activity pattern edges, enlongated contours, and whole textures ``seen'' in
the visual field. This representation has a distinct periodically
microstructured pattern: the pinwheel map. Neurons preferentially fire for
edges with a given orientation and the preference changes continuously as a
function of cortical location, except at singularities, the pinwheel centers,
where the all the different orientations meet \cite{ROJ90,BON91b}.

Iso-orientation domains form continuous bands or patches around pinwheels and,
on average, a region of about 1\,mm$^2$ (hypercolumn) will contain all 
possible orientation preferences. This topographical arrangement allows one
hypercolumn to analyze all orientations coming from a small area in the visual
field, but, as a consequence, the cortical representation of continuous contours
in the visual field would be depicted in a patchy, discontinuous fashion
\cite{EYS99}. In general, spatially separated elements are bound together by
short- and long-range lateral connections. While the strength of the local
short-range connection within one hypercolum is a graded function of cortical
distance, mostly independent of relative orientation \cite{DAS99}, long-range
connections over several hypercolums connect only iso-orientation domains of similar
orientation preference \cite{GIL92a,GIL96b}.
Even nearby regions, which are directly excitatory connected, have an
inhibitory component through local inhibitory interneurons and this is likely
be used to analyze angular visual features such as corners or T junctions
\cite{DAS99}.

Given the arguments above, we can now obtain localized yet spatially structured
activity states on the scale we aim for as initial conditions by using
iso-orientation domains that form continuous patches around pinwheels and
extend in a discontinuous fashion over larger areas.

In~\cite{NIE94c} the authors analyzed the design principles that lie behind the columnar organization of the visual cortex. 
The precise design principles of this cortical organization is governed by an annulus-like
spectral structure in Fourier domain \cite{ROJ90,NIE94c}, which is determined
by mainly one parameter (\emph{scaling}), that is, the annulus width. The
parameter \emph{depth} reflects the tuning properties of orientation preference
or we can also interpret this as the range of orientation angles that we
consider within the iso-orientation domain.  The third parameter reflects the
distance long-range coupling ranges before it significantly attenuates.

These design principles can be exploited and a procedure can be designed to construct maps with the same properties. The constructed maps come very close to the maps found in brains of macaque monkeys (see~\cite{NIE94c} and references therein).

To construct initial conditions from these maps we used a procedure that uses
four control parameters and is visualized in Fig.~\ref{fig:ini_con_schematic}.
The details are as follows: A pinwheel map is a function that maps our
twodimensional plane to the interval $(-\pi/2, \pi/2]$. We  construct such a
map using the procedure in \cite{NIE94c}. During construction, we can choose
the \emph{scaling} of the map. This is our first parameter.  After constructing
this map, by means of a gaussian, we choose a range of orientations that is
excited. Mathematically speaking this is the concatenation of the gaussian
distribution with the pinwheel map. This gives the next parameter, namely the
width of the gaussian that selects the angles, we call that parameter the
\emph{depth}. The next step is to constrain the generated pattern spatially by
multiplication with another gaussian which is defined on the plane $P$ and
chosen to be rotationally symmetric. The width of this gaussian gives rise to
the third parameter, the \emph{size} of the pattern. Finally we multiply the
pattern by a certain amplitude, which is chosen such that the integral of the
pattern over the plane gives a chosen number, which constitutes the fourth
parameter, we called the \emph{excess}.

Finally, initial conditions are generated by setting the plane to the (stable) homogeneous state and then adding the generated pattern to the activator variable $u$.

In a first run, we scanned the space spanned by the four parameters coarsely. We used the marginal distributions of the number of solutions with ED$>0$ with respect to the parameters to decide how densely to sample the parameter space in the final run.


\section*{Acknowledgments} The authors kindly acknowledge the support from the
Deutsche Forschungsgemeinschaft (DFG) in the frameworks of SFB910 and GRK1558, 
and from the Bundesministerium f\"ur Bildung und Forschung under grant BMBF 01GQ1109. 
MD has been supported in part also by the Mathematical Biosciences
Institute at the Ohio State University and the National Science Foundation
under Grant No. DMS 0931642. The authors would also like to thank Gerold Baier,
Michael Guevara, Zachary Kilpatrick, and Eckehard Sch\"oll for helpful
discussions and advice.

\section*{References}

%



\end{document}